%% file: main.tex
\documentclass[final]{llncs}

\input{preamble/macros.tex}

\input{preamble/packages.tex}
\input{preamble/mycolor.tex}
\input{preamble/style.tex}

\bibliography{citations}

\begin{document}\sloppy

\title{Performance-Portable Many-Core Plasma Simulations: Porting PIConGPU to OpenPower and Beyond
\thanks{This project has received funding from the European Union’s Horizon 2020 research and innovation programme under grant agreement No 654220}%
}

\author{Erik Zenker$^{1,2}$, Ren\'{e} Widera$^{1}$, Axel Huebl$^{1,2}$, Guido Juckeland$^{1}$, \\Andreas Kn\"{u}pfer$^{2}$, Wolfgang E. Nagel$^{2}$,  Michael Bussmann$^{1}$}
\institute{$^{1}$Helmholtz-Zentrum Dresden--Rossendorf, Dresden, Germany\\
  $^{2}$Technische Universit\"at Dresden, Dresden, Germany\\
  \email{\{e.zenker,r.widera,a.huebl,g.juckeland,m.bussmann\}@hzdr.de\\ \{andreas.knuepfer, wolfgang.nagel\}@tu-dresden.de}}

\maketitle

\begin{abstract}
  With the appearance of the heterogeneous platform OpenPower, many-core accelerator devices have been coupled with Power host processors for the first time. 
  Towards utilizing their full potential, it is worth investigating performance portable algorithms that allow to choose the best-fitting hardware for each domain-specific compute task.
  Suiting even the high level of parallelism on modern GPGPUs, our presented approach relies heavily on abstract meta-programming techniques, which are essential to focus on fine-grained tuning rather than code porting.
  With this in mind, the CUDA-based open-source plasma simulation code PIConGPU is currently being abstracted to support the heterogeneous OpenPower platform using our fast porting interface cupla, which wraps the abstract parallel C++11 kernel acceleration library Alpaka.

  We demonstrate how PIConGPU can benefit from the tunable kernel execution strategies of the Alpaka library, achieving portability and performance with single-source kernels on conventional CPUs, Power8 CPUs and NVIDIA GPUs.
\end{abstract}

  

\input{content/introduction}
\input{content/preliminaries}
\input{content/porting}
\input{content/evaluation}

\input{content/conclusion}

\newpage
\printbibliography

\end{document}

%% file: preamble/macros.tex
\newcommand{\blist}{\begin{easylist}[itemize]}
\newcommand{\elist}{\end{easylist}}
\newcommand{\alpaka}{Alpaka\xspace}
\newcommand{\picongpu}{PIConGPU\xspace}
\newcommand{\scfly}{SCFLY\xspace}
\newcommand{\cupla}{cupla\xspace}
\newcommand{\openpower}{OpenPower\xspace}
\newcommand{\cuda}{{CUDA}\xspace}

\newcommand{\openmp}{{OpenMP}\xspace}
\newcommand{\openacc}{{OpenACC}\xspace}
\newcommand{\opencl}{{OpenCL}\xspace}

\newcommand{\nvidia}{{NVIDIA}\xspace}
\newcommand{\volta}{{Volta}\xspace}
\newcommand{\intel}{{Intel}\xspace}
\newcommand{\amd}{{AMD}\xspace}
\newcommand{\ibm}{{IBM}\xspace}
\newcommand{\power}{{Power8}\xspace}
\newcommand{\powerNine}{{Power9}\xspace}

\newcommand{\cpp}[1]{\lstinline[identifierstyle=\color{black}\bfseries]{#1}}

\newcommand{\ignore}[1]{}

%% file: preamble/packages.tex
\usepackage[table]{xcolor}
\usepackage[
colorinlistoftodos,
textsize=scriptsize,
]{todonotes}
\usepackage[sharp]{easylist}
\usepackage{listings}
\usepackage{amsmath}
\usepackage{parcolumns}
\usepackage{pifont}
\usepackage{xspace}
\usepackage{amssymb}
\usepackage[style=numeric-comp, maxnames=99, backend=bibtex]{biblatex}
\usepackage{ tipa }

%% file: preamble/mycolor.tex
\definecolor{mygreen}{rgb}{0,0.6,0}
\definecolor{mygray}{rgb}{0.5,0.5,0.5}
\definecolor{mymauve}{rgb}{0.58,0,0.82}
\definecolor{myblue}{HTML}{E5F2FF}

%% file: preamble/style.tex
\lstdefinestyle{customptx}{
	numberstyle=\tiny\color{black},
	breaklines=true,
	basicstyle=\footnotesize\ttfamily,
	keywordstyle=\bfseries\color{blue}\ttfamily,
	stringstyle=\color{brown}\ttfamily,
	commentstyle=\itshape\color{mygreen}\ttfamily,
	belowcaptionskip=1\baselineskip,
	tabsize=4,
	xleftmargin=\parindent,
	language=TeX,
	showstringspaces=false,
	identifierstyle=\bfseries\color{myblue},
}

%% file: content/introduction.tex
\section{Introduction}
\label{introduction}
\picongpu~\cite{burau2010picongpu} is a fully-relativistic, multi-GPU, 3D3V particle-in-cell (PIC) code.
As such it allows to model the mutual interaction between electromagnetic fields and charged particles,
including effects of retardation in special relativity~(SRT) and the collective motion of collisionless plasmas, by solving Maxwell's equations self-consistently for charged particles and electromagnetic fields.
Besides the satisfied demand for large scales and high resolutions by computing the whole PIC cycle on GPUs, simulations of laser-ion acceleration from overdense targets~\cite{zeil2012direct} induce a further complexity in the dynamics of the plasma from collisional excitation and ionization processes.
As the free electron density from ionization processes determines intrinsic observables such as the plasma wavelength, the modeling of underlying quantum processes needs to be taken into account and is not yet covered in the plain electrodynamics provided by PIC.
Our approach to enhance the PIC algorithm is therefore to add a Monte Carlo step in the simulation with 0-D atom physics from \scfly~\cite{chung2007extension}. 
This method requires to calculate the transition rate matrix,
representing the likelihood of change of the atomic configuration of each ion from one time step to the next.
Each of the quantum processes has its own individual models, calibrated with experimental and theoretical estimates.
Even when considering the reduction of possible transitions by using an effective number of states, removing physically forbidden and very unlikely transitions, the total number of transitions can grow quadratically with the number of considered configurations.
In combination with the dependency of the transition matrix elements on local quantities, such as the energy distribution of neighboring electrons and photons of each individual ion in the plasma, the required amount of memory can easily grow into the size of several dozen gigabytes for a non-equilibrium system.

None of the accelerators that are currently available or announced for the near future fulfill these memory requirements. 
However, the accelerator's host system provides access to fast and large main memories and file systems.
The host's CPUs are used as a first computing stage to reduce the full transition matrix to smaller lookup tables.
CPUs excel at this task, since they typically provide better performance on trigonometric functions and implicit solvers.
Accordingly, only relevant data needs to be streamed to the GPU.

The \openpower platform couples various advanced hardware technologies on the same system~\cite{openpower} such as Power CPUs, \nvidia GPUs, and fast CPU--GPU interconnect technology~\cite{foley2014nvlink}.
To fully utilize the compute power
of this platform, it is currently necessary to use various programming models such as \cuda for GPU and \openmp for CPU.
However, this style of programming has the disadvantage that the code is difficult to maintain and it requires more work to switch algorithms between GPU and CPU implementations.
A uniform programming model allows to selectively determine the kernel execution hardware depending on the algorithmic requirements.
These requirements depend on the models of the individual physical process: some are memory bound, some compute bound, and the user chooses, based on domain knowledge and the relevance, on which hardware these processes should be executed.

Currently, widely utilized uniform parallel programming models such as OpenCL~\cite{stone2010opencl} do not fulfill all our requirements of a sustainable, open, maintainable, testable, optimizable, and single-source programming model.
Loop and container based approaches such as RAJA~\cite{hornung2014raja}, Kokkos~\cite{edwards2014kokkos}, and OpenMP 4.0~\cite{openmp2013openmp} would require a complete redesign of the \cuda based \picongpu code.
With \emph{\alpaka}~\cite{zewowi2016}, there exists an interface for parallel kernel acceleration which enables the programmer to compile single-source C++ kernels to various architectures, while providing all the requirements mentioned above.
As a first step to selective kernel acceleration on the \openpower platform, \picongpu has been ported with the \cuda-like interface \emph{\cupla}~\cite{cupla} to \alpaka, which currently allows for an execution either on the CPU or on the GPU.

This paper is structured as follows.
In Section~\ref{preliminaries}, we give a brief overview on \picongpu, \alpaka, and \cupla.
In Section~\ref{porting}, we provide our experiences on porting \picongpu with \cupla from \cuda to \alpaka.
Finally, the ported prototype is evaluated on various architectures in Section~\ref{evaluation}.

%% file: content/preliminaries.tex
\section{Preliminaries}
\label{preliminaries}



\subsection{\picongpu}

\picongpu is a multi-GPU particle-in-cell (PIC) code for three-dimensional field--particle interaction with high spatial resolution.
The code decomposes its global simulation domains into a grid of cells.
Cells are grouped into a cuboid volume called \emph{super cell}, and multiple of these super cells are again grouped into a cuboid volume which defines the local simulation domain of a single GPU.

Additionally, there is a second, spatially continuous domain for finite size macro particles such as ions and electrons.
They are able to move through the cells and interact with them, making PIC a particle mesh algorithm~\cite{bussmann2013radiative}.
Macro particles are grouped in frames, where each frame contains as many macro particles as there are cells in a super cell.
Frames are stored in a doubly linked list and correspond to a particular super cell.


Most of the operations on the cells are local stencils which include only a few neighboring cells and are therefore well suited to \cuda programming model of a multidimensional grid.
\picongpu is mapped to this model as follows:
The local simulation domain is mapped to the grid of a single GPU.
A super cell is mapped to a block that contains as many threads as there are cells --- in our simulation this amounts usually to 256 cells.
A thread calculates the field of a cell and its proportion of particles of its super cell.




\subsection{\alpaka and \cupla}
\label{alpaka}
\alpaka provides a uniform, abstract C++ interface to a range of parallel programming models.
It can express multiple levels of parallelism and allows for generic programming of kernels either for a single accelerator device or a single address space with multiple CPU cores.
The \alpaka abstraction of parallelization is influenced by and based on the groundbreaking \cuda abstraction of a multidimensional grid of blocks of threads.
The four main execution hierarchies introduced by \alpaka are called \emph{grid}, \emph{block}, \emph{thread}, and \emph{element} level.
The element level denotes an amount of work a thread needs to process sequentially.
These levels describe an index space which is called \emph{work division}.
Other programming models call these levels differently e.g. \opencl \emph{work-groups} of \emph{work-items}, \openmp \emph{teams} of \emph{threads}, and \openacc \emph{gang, worker, and vector}.

Separating parallelization abstraction from specific hardware capabilities allows for an explicit mapping of these levels to hardware.
The current implementation includes mappings to programming models, called back-ends, such as \openmp, \cuda, C++ threads, and boost fibers~\cite{fibers}.
Nevertheless, mapping implementations are not limited to these choices and can be extended or adapted for application-specific optimizations.
Which back-end and work division to utilize is parameterized per kernel within the user code.

A fast approach to port \cuda code to \alpaka is provided by the \cuda-like \alpaka interface \cupla $[q\chi ap\,'la?]$.
Cupla leaves most \cuda API calls unchanged, yet performs \alpaka calls in the background.
Thus, \cupla provides a simple and fast porting approach by reducing the number of lines  of the original \cuda code a programmer needs to modify.



%% file: content/porting.tex
\section{Porting with \cupla}
\label{porting}

In this section we discuss the steps necessary to port the \cuda accelerated code of \picongpu from GPU to CPU hardware.
Our approach is to replace \cuda by the \cuda-like interface \cupla.
Afterwards, we can utilize Alpaka's \cuda and {\openmp}~2.0 back-ends to execute our kernels on both GPUs and CPUs.  



Cupla leaves most parts of the \cuda code unchanged such as memory allocations, memory copies, stream handling, device handling, and index queries.
The programmer is still required to handle three porting steps.
Firstly, the \texttt{cuda\_runtime.hpp} include has to be replaced by \texttt{cuda\_to\_cupla.hpp} and all \texttt{.cu} files renamed to \texttt{.cpp}.
Secondly, The \texttt{\_\_host\_\_}, \texttt{\_\_device\_\_}, and \texttt{\_\_global\_\_} keywords need to be replaced by equivalent \cupla macros and \cuda global functions rewritten into parenthesis operators of C++ functors.
The accelerator object of the accelerator template type has to be passed to these operators and the underlying device functions.
Finally, each shared memory allocation has to be replaced by an equivalent \cupla macro.
Listing~\ref{fig:port_snippet} shows equivalent \cuda and \cupla code snippets of a kernel function initializing an array by a constant value. 
In contrast to the \cuda kernel, each thread of the \cupla kernel loops over the \texttt{x} dimension of the element level.

\begin{figure}[!t]
\begin{lstlisting}
// CUDA Kernel
__global__ void kernel ( int * data )
{
    int id = blockDim.x * blockIdx.x 
             + threadIdx.x;
    data[ id ] = 42;
}
\end{lstlisting}
\begin{lstlisting}
// Alpaka Kernel
struct void kernel {
  template < typename Acc >
  ALPAKA_FN_ACC void operator () (
    Acc const & acc,
    int * data
  ) const 
  { 
      int id = blockDim.x * blockIdx.x * elemDim.x 
               + threadIdx.x * elemDim.x;
      for( int elem = 0; elem < elemDim.x; ++elem)
        data[ id + elem ] = 42
  }
};
\end{lstlisting}
\caption{\cuda and \cupla kernels which initialize each element in the input array \cpp{data} by the value 42. The \cupla kernel on the bottom was created through wrapping the \cuda kernel on the top within a C++ functor. Each thread of the \cupla kernel processes multiple elements through looping over the dimensions of the additional element level. In the \cupla kernel \texttt{blockDim}, \texttt{blockIdx}, \texttt{threadIdx}, and \texttt{elemDim} are pre-processor macros accessing the \texttt{acc} variable.}
\label{fig:port_snippet} 
\end{figure}


The native PIC code
consists of about forty~thousand lines of code.
This code is a mixture C++11 and platform-dependent \cuda code.
R. Widera programmed about two days, applied the \cupla porting steps mentioned above, touched most of our nine hundred device functions, forty kernels, amounting to two~thousand lines of code, to provide the first \alpaka based prototype.
Although this prototype did not utilize the \emph{element} level, it was already executable on both a \power device using the {\openmp}~2.0 back-end and on an \nvidia device using the \cuda back-end.
The number of threads in a block was left unchanged.
Accordingly, the domain of a super cell is processed by a block consisting of 256 threads.

This block-size leads to inefficient communication between threads on the \power when the the element level is omitted, resulting in more frequent cache misses and a decrease in performance.
Accordingly, the integration of the element level enables for a work division of blocks with a single thread and multiple elements to calculate the entire domain of a super cell.
This provides a more efficient mapping of \alpaka-threads to hardware threads and, therefore, an improved vectorization and cache utilization by the compiler.
The integration of the element level required to loop over the fixed-size element index space for each sequential kernel part.
These sequential parts were wrapped in lambda functions.
Furthermore, single element variables were expanded to multidimensional fixed-size arrays.
This change, on three thousand lines of code, took our developer about ten days.

To sum up, our developer modified about five thousand lines of code in a matter of two weeks, after which the entire forty thousand lines \picongpu code could be compiled and run efficiently on CPU and GPU devices.
It was not necessary to modify the core data structures or algorithms of \picongpu.
The element level has been added to enable a single thread to process the domain of a super cell.
In the following section we will evaluate the performance of our {\alpaka}-based PIC simulation on both architectures.



%% file: content/evaluation.tex
\section{Evaluation}
\label{evaluation}

This section provides the evaluation of the \picongpu code~\cite{rene_widera_2016_53761} that was ported to various compute architectures
(see Table~\ref{tab:hardware}).
We measured the runtime and performance
of the memory-bound PIC algorithm as implemented in \picongpu with a simulation of the Kelvin-Helmholtz instability~\cite{bussmann2013radiative} for one thousand time steps in double and single precision and compared these results between the various architectures.
The simulation was parameterized with the Boris pusher, Esirkepov current solver, Yee field solver, trilinear interpolation in field gathering, three spatial dimensions~(3D3V), 128~cells in each dimension, electron and ion species with each sixteen particles per cell, and quadratic-spline interpolation (TSC)~\cite{hockney1988computer}. 
On all CPU devices the {\openmp}~2.0 back-end was used with a block consisting of a single thread with 256 elements.
On \nvidia GPUs the \cuda back-end is used with a block consisting of 256 threads with a single element.
All GPU evaluations are compiled with nvcc\footnote{\texttt{-{}-use\_fast\_math -{}-ftz=false -g0 -O3 -m64}} 7.0 and all CPU evaluations with gcc\footnote{\texttt{-g0 -O3 -m64 -funroll-loops -march=native -{}-param max-unroll-times=512 \mbox{-ffast-math}}} 4.9.2.

\begin{table*}[!t]
  \scriptsize
  \begin{center}
    \caption{Compute nodes for evaluation (core counts in braces are HW threads).}
    \begin{tabular}{ |l || c | c | c |  c | c | c |}
      \hline
      \textbf{Vendor}                       & \amd           & \intel              & \ibm                  & \nvidia\\
      \hline
      \textbf{Architecture}                 & Interlagos~\cite{amdOpteron}     & Haswell~\cite{xeonE5}             & Power8~\cite{fluhr20145}                & Kepler~\cite{k80}\\
      \hline       
      \textbf{Model}                        & Opteron 6276   & Xeon E5-2698v3      & Power8 8247-42L       & K80 GK210\\
      \hline
      \textbf{Used devices per node}        & 4              & 2                   & 2                     & 1\\
      \hline
      \textbf{Cores per device}             & 16             & 16 (32)             & 10 (80)               & 2496\\
      \hline
      \textbf{Base clock frequency}         & 2.3 GHz        & 2.3  GHz            & 2.1 GHz               & 0.56  GHz\\
      \hline
      \textbf{Release date}                 & Q4/2011        & Q3/2014             & Q1/2014               & Q4/2014\\
      \hline
      \textbf{Peak performance(sp)}         & 960 GFLOPS     & 2354 GFLOPS         & 1120 GFLOPS           & 4350 GFLOPS\\
      \hline      
      \textbf{Peak performance(dp)}         & 480 GFLOPS     & 1177 GFLOPS         & 560 GFLOPS            & 1450 GFLOPS\\
      \hline
    \end{tabular}
    \label{tab:hardware}
  \end{center}
  \vspace{-3em}
\end{table*}

\begin{figure}[b]
  \centerline
      {\resizebox{\textwidth}{!}{\includegraphics{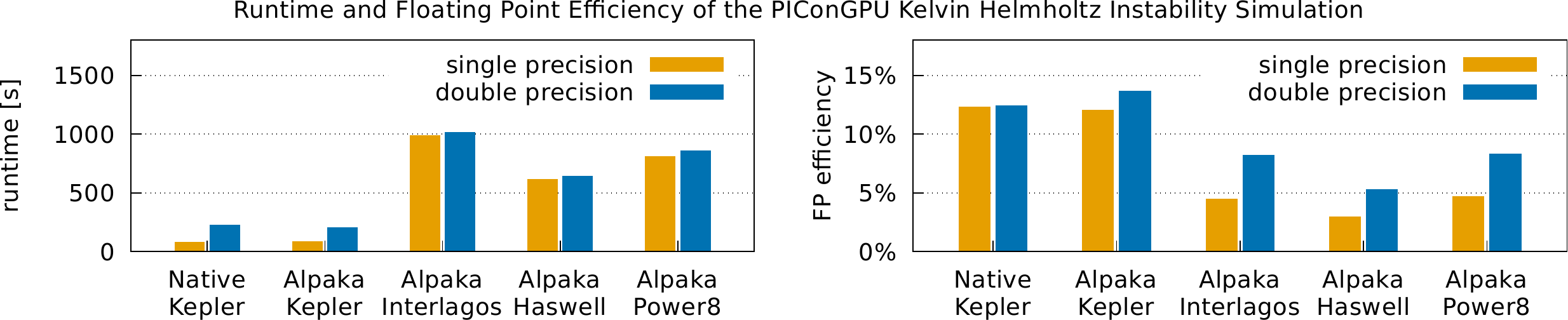}}}
      \caption{As an example to evaluate a memory-bound PIC code, runtime and floating point efficiency of the \picongpu Kelvin-Helmholtz instability simulation for single precision and for double precision was measured on various architectures (see Table~\ref{tab:hardware}).}
          \label{fig:multiplot}
          \vspace{-1em}
\end{figure}

Figure~\ref{fig:multiplot} displays the measured runtime and efficiency of the evaluated simulation.
On the \nvidia K80, the differences in runtime between the native and the ported PIC code are about one percent for single precision. 
For double precision, the \alpaka based code is even faster, because \alpaka emulates double \emph{atomicAdd} using \emph{atomicCAS} instead of the slower \emph{atomicExch} used by the native \picongpu implementation.
Nevertheless, this small optimization could have been introduced easily into the native \picongpu code to achieve the same runtime results.
According to these measurements, \alpaka can keep its promise of zero-overhead abstraction on the same architecture even for rather complex applications such as \picongpu.
The runtime between GPU and CPU implementations differ in one order of magnitude for single precision.
However, the results need to be evaluated in relation to the theoretical peak performance of the particular architecture.
This metric is denoted as floating point efficiency in Figure~\ref{fig:multiplot}.
Regarding floating point efficiency, CPU and GPU vary by a factor of three to four on single precision and by a factor of two on double precision.
Thus, \alpaka provides not just portability between GPU and CPU, but decent performance on both.
All evaluated CPU architectures show similar runtime and efficiency characteristics.  
Nevertheless, the \intel architecture offers the lowest runtime and highest (theoretical) peak performance of all evaluated CPU devices.
However, there still exists some potential to increase performance, as it only provides five percent floating point efficiency on double precision.
While the IBM and AMD architectures fare slightly better with about eight percent double precision efficiency, there is still a lot of potential compared to the GPU efficiency.
By refining the \alpaka back-ends and tuning the work division, this potential can be utilized to increase the performance of the CPU architectures even more.

%% file: content/conclusion.tex
\section{Conclusion}
We have presented the current progress in porting the particle-in-cell simulation \picongpu onto the \openpower platform through utilizing the \cuda-like \alpaka interface \emph{\cupla}.
The core routines of the forty thousand lines mixed C++ and \cuda code have been ported from \cuda to \alpaka within two weeks.
Through this abstraction, the ported \picongpu implementation is executable on \amd, \ibm, \intel, and \nvidia architectures.
The code was not just ported, but has been moved to a generic \emph{single-source} multi-platform programming model.
Thus, \picongpu never needs to be ported again.

The native \cuda version and the \alpaka version show no significant differences in runtime or performance on the \nvidia hardware, which demonstrates zero overhead abstraction capabilities of \alpaka.
GPU and CPU devices differ in a factor of about two in efficiency on double precision, providing decent performance among the evaluated architectures.  

Future work will focus on the evaluation of each kernel on CPU and GPU hardware separately.
Based on these measurements, we want to provide a static mapping of kernels to heterogeneous hardware to achieve the best possible overall performance on the particular HPC system.
Furthermore, we want to complete the porting of the remaining simulation plugins within \picongpu and add a more fine-grain element level implementation.

The code is ready for the upcoming \powerNine and \nvidia~\volta-based heterogeneous systems such as Summit~\cite{summit} at the Oak Ridge National Laboratory.
By using \alpaka we have the possibility to optimize and adapt our back-ends to these systems once they are fully specified and available for evaluation.